\documentclass[12pt]{article}
\usepackage{graphicx,amsfonts,amsmath,amssymb,mathrsfs,url,verbatim}

\title{Many-Worlds\\ and Schr\"odinger's First Quantum Theory}
\author{
Valia Allori\footnote{Department of Philosophy, Northern Illinois University,
     Zulauf Hall 920, DeKalb, IL 60115, USA. E-mail: vallori@niu.edu},
Sheldon Goldstein\footnote{Departments of Mathematics, Physics and
     Philosophy, Hill Center, Rutgers, The State University of New  
     Jersey, 110 Frelinghuysen Road, Piscataway, NJ 08854-8019, USA.
     E-mail: oldstein@math.rutgers.edu},\\
Roderich Tumulka\footnote{Department of Mathematics,
     Rutgers University, Hill Center,  
     110 Frelinghuysen Road, Piscataway, NJ 08854-8019, USA.
     E-mail: tumulka@math.rutgers.edu},
 and Nino Zangh\`\i\footnote{Dipartimento di Fisica dell'Universit\`a
     di Genova and INFN sezione di Genova, Via Dodecaneso 33, 16146
     Genova, Italy. E-mail: zanghi@ge.infn.it}
}
\date{August 5, 2009}
\newcommand{\Hilbert}{\mathscr{H}}

\renewcommand{\Im}{\mathrm{Im}}

\newcommand{\RRR}{\mathbb{R}}

\newcommand{\scp}[2]{\langle #1|#2 \rangle}

\newcommand{\tr}{\mathrm{tr}}
\newcommand{\num}{n}
\newcommand{\Lnum}{\mathscr{L}}
\newcommand{\Lset}{L}

\newcommand{\n}[1]{{#1}}
\newcommand{\x}[1]{{#1}}
\newcommand{\z}[1]{{#1}}

\begin{document}
\maketitle%\sloppy
\begin{abstract}
Schr\"odinger's first proposal for the interpretation of quantum mechanics was based on a postulate relating the wave function on configuration space to  charge density  in physical space. Schr\"odinger apparently later thought that his proposal was empirically wrong. We argue here that this is not the case, at least for a very similar proposal with charge density replaced by mass density. We argue that when analyzed carefully this theory is seen to be an empirically adequate many-worlds theory and not an empirically inadequate theory describing a single world. Moreover, this formulation---Schr\"odinger's first quantum theory---can be regarded as a formulation of the many-worlds view of quantum mechanics that is ontologically clearer than Everett's.
\medskip

\end{abstract}

%\tableofcontents

\newpage
\section{Monstrosity}

\begin{quotation}
\noindent\textit{The `many world interpretation' \ldots\ may have something distinctive to say in connection with the `Einstein Podolsky Rosen puzzle', and it would be worthwhile, I think, to formulate some precise version of it to see if this is really so.}\hfill John S.\ Bell \cite{Bell86b}
\end{quotation}
The many-worlds view of quantum mechanics is  popular, but also controversial; it is very radical and eccentric, but also inspiring. It is an incarnation of the desire to abolish the vague division of the world, introduced by the Copenhagen interpretation, into system and observer, or quantum and classical, and to obtain a fully precise formulation of quantum mechanics, in which the axioms do not concern observers and observation but reality. In the words of Hugh Everett \cite{Eve57b}, the inventor of the many-worlds view:

\begin{quotation}
The Copenhagen Interpretation is hopelessly incomplete because of its a priori reliance on classical physics \ldots\ as well as a philosophic monstrosity with a ``reality'' concept for the macroscopic world and denial of the same for the microcosm.
\end{quotation}

We report here on some considerations on the many-worlds view of quantum mechanics inspired by \x{Erwin} Schr\"odinger's \cite{sch1} original interpretation of the wave function $\psi$ on configuration space as generating a continuous distribution of matter (or charge) spread out in physical space. 
As we shall explain, Schr\"odinger's original version of quantum mechanics may be regarded as a version of many-worlds---though some adherents of many-worlds will presumably not regard it as such---that we think is worth considering. It is a version that, in our opinion, qualifies as a ``precise version of'' many-worlds such as Bell called for in the passage quoted above.

\section{Duality}

Let us describe Schr\"odinger's first quantum theory in our own words.
Think first about classical mechanics. Matter consists of particles, moving along trajectories defined by the equations of the theory. Alternatively, a classical theory could claim that instead of consisting of particles, matter is continuously distributed in 3-space and mathematically described by a function $m(x,t)$, where $x$ runs through physical 3-space, providing the spatial density of matter at time $t$. We call this ontology the \emph{matter density ontology}. Such a theory would involve classical equations governing the $m$ function. In the $m$ function we can find the macroscopic objects of our experience, such as tables and chairs, by noting that at a certain time there is a region of space, with the shape of a table or chair, in which the matter density is significantly higher than in the surroundings. In such a theory, it would be wrong to say that matter consists of a large number (such as $10^{23})$ of particles, since there are no particles in the ontology, just a continuum of stuff.

Now combine the matter density ontology with non-classical equations. Specifically, suppose that matter is continuously distributed with density $m(x,t)$, but now suppose that the $m$ function is given by the following equation:
\begin{equation}\label{mdef}
  m(x,t) = \sum_{i=1}^N m_i \int d^3x_1 \cdots d^3x_N \, \delta^3(x-x_i) \, 
  \bigl|\psi_t(x_1,\ldots, x_N) \bigr|^2\,.
\end{equation}
Here, $\psi_t$ is a wave function as in quantum mechanics, a function on $\RRR^{3N}$ evolving according to the usual Schr\"odinger equation
\begin{equation}\label{Schr}
  i\hbar \frac{\partial \psi}{\partial t} = 
  - \sum_{i=1}^N \frac{\hbar^2}{2m_i} \nabla_i^2 \psi + V \psi\,,
\end{equation}
and $m_i$ denotes the mass of particle $i$, $i= 1, \ldots, N$.

The $m$ function \eqref{mdef} is basically the natural density function in 3-space that one can obtain from the $|\psi|^2$ distribution in configuration space. The formula means that, starting from $|\psi|^2$, one integrates out the positions of $N-1$ particles to obtain a density in 3-space. Since the number $i$ of the particle that was not integrated out is arbitrary, it gets averaged over. The weights $m_i$ are the masses associated with the variables $x_i$, which may seem the most natural choice for defining the density of matter.

This provides, in fact, already the complete specification of a physical theory. In the terminology of \cite{AGTZ06}, this theory is called  ``Sm'' (S for the Schr\"odinger equation and m for the $m$ function). It is closely related to---if not precisely the same as---the version of quantum mechanics first proposed by Schr\"odinger \cite{sch1}. 
After all, Schr\"odinger originally regarded his theory as describing a
continuous distribution of matter (or charge) spread out in {\em physical
space} in accord with the wave function on {\em configuration space}
\cite[p.~120]{sch}:
 \begin{quotation}
 We had calculated the density of electricity at an arbitrary point in
 space as follows. We selected \emph{one} particle, kept the trio of
 co-ordinates that describes \emph{its} position in ordinary mechanics
 fixed; integrated $\psi\overline{\psi}$ over all the rest of the
 co-ordinates of the system and multiplied the result by a certain
 constant, the ``charge'' of the selected particle; we did a similar thing
 for each particle (trio of co-ordinates), in each case giving the
 selected particle the same position, namely, the position of the point of
 \emph{space} at which we desired to know the electric density. The latter
 is equal to the algebraic sum of the partial results.
 \end{quotation}
This is just a verbal description of the formula \eqref{mdef}, except with charges instead of masses.\footnote{If we replace the masses $m_i$ in \eqref{mdef} with the charges $e_i$, as Schr\"odinger did, then the following problem arises that is absent when using masses. If the wave function of a macroscopic body (say, a piece of wood) is such that the Heisenberg position uncertainties of the atomic nuclei are of the order of an Angstrom, i.e., of the order of the size of an atom, then the positive charge of a nucleus may be smeared out over the same volume as the negative charge of the electrons, so that they may cancel each other, leaving only a negligible remainder in the $m$ function. In this case, the macroscopic body would hardly be recognizable in the $m$ function, and such an $m$ function would not provide a plausible image of our world. \z{This problem notwithstanding}, replacing the masses $m_i$ in \eqref{mdef} with the charges $e_i$, or with the constant value 1, leads to theories which are empirically equivalent to Sm and similar to Sm in all relevant \z{respects}. In particular, our conclusions about nonlocality (Section \ref{sec:locality}) and probability (Section \ref{sec:probability}) for Sm apply equally to these \z{theories}.}
Schr\"odinger soon rejected this theory because he thought that it rather clearly conflicted with experiment. After all, the spreading of the matter density arising from equation \eqref{Schr} would appear to contradict the familiar localized detection events for quantum particles, such as in the two-slit experiment. Moreover, given that there are no particles in Sm, but instead matter is really continuous, one might think at first that Sm is empirically refuted by the evidence for the existence of atoms. Yet, Schr\"odinger's rejection was perhaps a bit hasty, as we will see. Be that as it may, Schr\"odinger did in fact create the first many-worlds theory, though he probably was not aware that he had done so.

It is easy to see that Sm has a certain many-worlds character, since if $\psi$ is the wave function of Schr\"odinger's cat then there will be two contributions to the $m$ function, one resembling a dead cat and the other a live cat. We will say more about this in Section~\ref{sec:parallelity}. For now note the duality: there exist two things, the wave function $\psi$ and the matter density function $m$. The latter represents the ``primitive ontology'' (PO) of the theory \cite{AGTZ06}, the \z{elements of} the theoretical picture that correspond to matter in 3-dimensional space; the wave function tells the matter how to move. The notion of PO is closely connected with what Bell called the ``local beables'':
\begin{quote}
  [I]n the words of Bohr, `it is decisive to recognize that, however
  far the phenomena transcend the scope of classical physical
  explanation, the account of all evidence must be expressed in
  classical terms'. It is the ambition of the theory of local beables
  to bring these `classical terms' into the equations, and not
  relegate them entirely to the surrounding talk.\hfill \cite{Bell76}
\end{quote}
  \z{We note that the matter density $m(x,t)$ \eqref{mdef}, defined as it is on physical space, is given by local beables, while the wave function $\psi = \psi(x_1,\ldots, x_N)$, defined on configuration space, is not.}

To introduce a PO for a theory means to be explicit about what space-time entities the theory is fundamentally about. There are various possibilities for what type of mathematical objects could represent the elements of the PO, including particle world lines as in classical or Bohmian mechanics, world sheets as maybe suggested by string theory, world points as in the GRW theory with the flash ontology \cite{Bell87,Tum04,AGTZ06}, or, instead of subsets of space-time, functions on space-time representing a field or a continuous density of matter, as in the matter-density ontology of Sm (and GRWm \cite{AGTZ06}). The wave function also belongs to the ontology of Sm, but not to the PO:  physical objects \z{in Sm} are made of $m$, not of $\psi$. \z{Rather the role of $\psi$ in this theory lies in the relation defined by \eqref{mdef} between $\psi$ and $m$. (That $m$ is primitive and $\psi$ is not should not be taken to imply that, contrary to \eqref{mdef},  $\psi$ should be defined in terms of $m$.)}

\[***\]

Let us compare Sm to Bohmian mechanics \cite{Bohm52,Bell66,Gol98}. The latter is a theory of particles with trajectories $Q_i(t) \in \RRR^3$, guided by a wave function $\psi$. As in classical mechanics, particles are points moving around in space, but the equation of motion is highly non-classical. In this theory there is a wave--particle duality in the literal sense: there is a wave, and there are particles. For understanding Bohmian mechanics it is important to think of these two parts of reality, $\psi$ and the $Q_i$, in a particular way: When one says, for example, that the pointer of an apparatus points to the value $\alpha$ then one means that the particles of which the pointer consists are at the appropriate positions corresponding to $\alpha$, but one does not mean that the wave function lies in the subspace of Hilbert space that can be associated with the description that the pointer is pointing to $\alpha$. To put this succinctly, one could say that the matter in Bohmian mechanics consists of the particles, not of the wave function. The role of the wave function, in contrast, is to tell the particles how to move. Indeed, the wave function $\psi$ occurs in the equation of motion for the particles,
\begin{equation}\label{Bohm}
  \frac{dQ_i}{dt} = \frac{\hbar}{m_i} \Im \frac{\psi^*\nabla_i \psi}{\psi^*\psi}(Q_1(t), \ldots, Q_N(t))\,.
\end{equation}
Here, the wave function $\psi=\psi_t$ evolves according to the Schr\"odinger equation \eqref{Schr}. It is consistent with these two equations that the configuration $Q(t) = (Q_1(t), \ldots, Q_N(t))$ has probability distribution $|\psi_t|^2$ at every time $t$.

Bohmian mechanics thus has the duality in common with Sm: In both theories, there are mathematical variables specifying the distribution of matter in 3-dimensional space---and not in $3N$-dimensional configuration space. Bohmian mechanics specifies this distribution by means of the actual configuration $Q(t) = (Q_1(t), \ldots, Q_N(t))$, and Sm of course by the $m(\cdot,t)$ function. A difference between Bohmian mechanics and Sm is that the $m$ function is a function of the quantum state $\psi$, whereas $Q$ is not. Indeed, in the initial value problem of Bohmian mechanics, we have to choose an initial value for $Q$ in addition to the initial value of $\psi$.

\[***\]

The many-worlds view is often presented as asserting that there exists only the wave function, which evolves unitarily, and nothing else. Let us call this view S0, according to a notation pattern that indicates first how the wave function evolves (Schr\"odinger equation) and then what the PO is (nothing).  
%We find it difficult to regard S0 as a meaningful physical theory. S0 seems to us to be not ``many worlds,'' but ``zero worlds.'' That is because S0 insists that there is no PO, corresponding to matter in space-time, which we feel means there is no matter, and thus no physics. (For additional discussion, see \cite{Vthesis}.)
\z{We} believe it is useful to clearly distinguish between S0 and Sm. Doing so affords a clear separation of the main issues for a many-worlds theory: the issue of whether a theory, in order to make clear sense as a physical theory, needs to posit a PO in space and time from the issues of whether the existence of parallel worlds is scientifically plausible, of whether the Bell inequality can be violated by a local theory, and of whether such a theory can give rise to the appearance of randomness.

\[***\]

We have defined Sm using the Schr\"odinger picture, but it can be formulated as well in the Heisenberg picture. To this end let
\begin{equation}\label{Mdef}
M(x) = \sum_{i=1}^N m_i \, \delta^3(x-\hat{Q}_i)
\end{equation}
be the mass density operator at $x\in\RRR^3$, with $\hat{Q}_i$ the triple of position operators associated with the $i$-th particle. Then \eqref{mdef} can be rewritten as
\begin{equation}
m(x,t) = \scp{\psi_t}{M(x)|\psi_t}\,,
\end{equation}
and this expression can be transferred to the Heisenberg picture in the usual way by setting
\begin{equation}
M(x,t) = \exp(iHt/\hbar) \, M(x) \, \exp(-iHt/\hbar)\,,
\end{equation}
so that
\begin{equation}
m(x,t) = \scp{\psi}{M(x,t)|\psi}\,.
\end{equation}
However, it will be convenient for us to continue using the Schr\"odinger picture.

\section{Parallelity} \label{sec:parallelity}

In Sm, apparatus pointers never point in a specific direction (except when a certain direction in orthodox quantum theory would have probability more or less one), but rather all directions are, so to speak, realized at once.  As a consequence, it would seem that its predictions do not agree with those of the quantum formalism. Still, it can be argued that Sm does not predict any \emph{observable} deviation from the quantum formalism: there is, arguably, no conceivable experiment that could help us decide whether our world is governed by Sm on the one hand or by the quantum formalism on the other.  Let us explain. 

Whenever the wave function (as a function on configuration space!)\ consists of disjoint packets $\psi_1,\ldots,\psi_\Lnum$,
\begin{equation}\label{psiell}
\psi = \sum_{\ell=1}^\Lnum \psi_\ell\,,
\end{equation}
it follows that
\begin{equation}\label{mmell}
  m(x) = \sum_{\ell=1}^\Lnum m_\ell(x)\,,
\end{equation}
where $m_\ell(x)$ is defined in terms of $\psi_\ell$ in the same way as $m(x)$ in terms of $\psi$ by \eqref{mdef}. Suppose further, as we shall henceforth do, that in particular the $\psi_\ell$ represent macroscopically different states, as with Schr\"odinger's cat. Then it is plausible that also in the future the $\psi_\ell$ will remain (approximately) disjoint (until Poincar\'e recurrence times), so that
\begin{equation}\label{mmellt}
  m(x,t) = \sum_{\ell=1}^\Lnum m_\ell(x,t)\,,
\end{equation}
with $m_\ell(\cdot,t)$ defined in terms of $\psi_{\ell,t}$ (the time-evolved $\psi_\ell$), also for $t$ in the future. Moreover, as long as $\psi_\ell$ does not itself become a superposition of macroscopically different states, $m_\ell$ behaves as expected of the macro-state of $\psi_\ell$ and provides a reasonable and recognizable story. 

For example, for Schr\"odinger's cat we have that $\psi=\psi_1 +\psi_2$ with $\psi_1$ the wave function of a live cat and $\psi_2$ the wave function of a dead cat, and $m_1(x,t)$ behaves like the mass density of a live cat (up to an overall factor), while $m_2(x,t)$ behaves like that of a dead cat. Note that, by the linearity of the Schr\"odinger evolution, the live cat and the dead cat, that is $m_1$ and $m_2$, do not interact with each other, as they correspond to $\psi_1$ and $\psi_2$, which would in the usual quantum theory be regarded as alternative states of the cat. The two cats are, so to speak, reciprocally transparent.

%To sum up, if one considers the matter densities \eqref{mdef} that correspond to macroscopic superpositions, one sees that they 
\z{More generally, consider an (evolving) decomposition \eqref{psiell} associated with an orthogonal decomposition  $\Hilbert=\oplus_\ell \Hilbert_\ell$ of the Hilbert space $\Hilbert$ into subspaces $\Hilbert_\ell$ corresponding to different macrostates  \cite{vN}. Then the components of the corresponding decomposition \eqref{mmell} should}
form independent families of correlated matter density associated with the terms of the superposition, with no interaction between the families. The families can indeed be regarded as comprising many parallel worlds, superimposed on a single space-time.  Metaphorically speaking, the universe according to Sm resembles the situation of a TV set that is not correctly tuned, so that one always sees a mixture of several channels. In principle, one might watch several movies at the same time in this way, with each movie conveying its own story composed of temporally and spatially correlated events. Thus, in Sm reality is very different from what we usually believe it  to be like. It is populated with ghosts we do not perceive, or rather, with what are like ghosts from our perspective, because the ghosts are as real as we are, and from their perspective we are the ghosts. Put differently, within the one universe consisting of matter with distribution $m(\cdot,t)$ in one space-time, there exist parallel worlds, many of which include separate, somehow different copies of the same person.

So the ``many worlds'' here are the many contributions $m_\ell$, and $\Lnum$ is the number of the different worlds. It is important to realize that the concept of a ``world'' does not enter in the definition of the theory, which consists merely of the postulate that $m(x,t)$ means the density of matter together with the laws \eqref{Schr} and \eqref{mdef} for $\psi$ and $m$. Instead, the concept of a ``world'' is just a practical matter, relevant to comparing the $m$ function provided by the theory to our observations, that may well remain a bit vague. There is no need for a precise definition of ``world,'' just as we can get along without a precise definition of ``table.''

While Sm has much in common with Everett's many-worlds formulation of quantum mechanics \cite{Eve57}, there are some differences. In Sm, the ``worlds'' are explicitly realized in the same space-time. Moreover, Sm has a clear PO upon which the existence and behavior of the macroscopic counterparts of our experience can be grounded. \z{Thus the ``preferred basis problem'' does not arise for Sm.} \z{Everett}'s view is essentially S0, as his worlds are thought of as corresponding directly to the various parts $\psi_\ell$ of the wave function, with no intervening matter densities $m_\ell$.

Since in Sm the wave function evolves according to the Schr\"odinger equation, it never collapses. Let us make explicit that this is not in conflict with the collapse rule of the quantum formalism (i.e., the algorithm for computing the statistics of outcomes of quantum experiments) because the formalism talks about wave functions of quantum objects whereas the $\psi$ in the defining equations \eqref{Schr} and \eqref{mdef} is really a wave function of the universe. In the quantum formalism, it seems meaningless to talk about a wave function of the universe since the wave function of a system is only used for statistical predictions of what an observer outside the system will see. In Sm, in contrast, the wave function of the universe is not meaningless at all, as it governs the behavior of the matter. 

As a consequence of the relation between the $m_\ell$ and the $\psi_\ell$, each world $m_\ell$ looks macroscopically like what most physicists would expect a world with wave function $\psi_\ell$ to look like macroscopically. This fact makes clear not only that tables and chairs can be found in $m_\ell$ but also that the possible outcomes of experiments are the same as in quantum mechanics; for example, particle detectors can only have integer numbers of clicks. In particular, the empirical evidence for the granular structure of matter (e.g., the existence of atoms, or the fact that electrons can be counted) is not in logical contradiction with the continuous nature of matter as postulated in Sm.

%\[***\]

\z{Readers may worry that the following \z{problem} arises in Sm. Since with every \x{non-deterministic} ``quantum measurement,'' each world splits into \x{several}, the number of worlds should increase exponentially with time. After adding very many contributions $m_\ell$, we may expect that $m$ looks like random noise, or like mush. The worry is that the separate stories corresponding to the $m_\ell$ then cannot be extracted any more from an analysis of $m$. However, when \x{we consider,} not just the $m$ function associated with the
present time, but also that in the past and in the future, then the
reasonable possibilities of splitting $m$ into causally disconnected,
branching, recognizable worlds $m_\ell$ are presumably very limited, and
\x{should more or less correspond to a splitting \eqref{psiell} of the wave
function} \z{based on an orthogonal decomposition of $\Hilbert$ into macrostates.}\footnote{\z{Appeal to causal disconnection and branching in the extraction of ``worlds'' from the quantum state and its image on physical space has been discussed in the contemporary literature on the Everett interpretation; see, e.g., \cite{saunders95, wallace03}.}}
 Thus, while it would be a problem for Sm if $m(x,t)$ were constant
as a function of $x$ and $t$, no problem \x{need arise} if $m(x,t)$ is highly \x{intricate}.}

\[***\]

We  \z{wish to address} a question that is often raised against the many-worlds view: If a conscious observer is in a superposition of very different brain states (say, having read the figure ``1'' and having read the figure ``2''), what is her or his conscious experience like? Sm entails that there are two persons, i.e., two contributions to the $m$ field, one behaving like a person who has read ``1'' and the other like a person who has read ``2''. So far so good, but that is only a statement about the behavior of matter, and does not strictly imply anything about the conscious experience. Since we cannot solve the mind--body problem, or get to the bottom of the nature of consciousness, we invoke an hypothesis of the kind that has always been implicitly used in physics, in particular in classical physics: the assumption of a suitable psycho-physical parallelism implying that  a person has a conscious experience  of the figure ``1'' whenever the person, more precisely the person's matter, is configured appropriately.

\section{Reality}\label{sec:reality}

In Sm, the right way to understand the theory is to regard the $m$ function as the basic reality, and not $\psi$. 
%When calling the $m$ function the ``primitive ontology,'' we do not mean that the wave function is not real, but rather that it exists in a different way, just as natural laws exist but not in the same way as matter, or space exists but not in the same way as matter. 
\z{The} way Sm connects with the world of our experiences is analogous to the way that  Bohmian mechanics does. There, the connection is made through the particles, not through the wave function.
\z{Insofar as a universe governed by Sm is concerned, the essential nature of the wave function is defined by its evolution and its relation to the $m$ function.}

\z{Sm, but not S0, requires that the causally disconnected entities which constitute ÒworldsÓ are part of or are \z{realized} in some \z{precisely-defined}, locally specifiable, spatio-temporal entity of a relatively familiar kind (the PO). And on this we disagree with contemporary advocates of the Everett interpretation such as Simon Saunders, Hilary Greaves, Max Tegmark, David Deutsch and David Wallace: we require, and they do not, that worlds be instantiated in such a way. And this  corresponds in turn to a disagreement about whether anything like a PO is required in a physical theory. }

%Now let us turn to the question: What is wrong with S0, which involves just the wave function and nothing else? Why does one need a PO at all? 
\z{We feel the need for a PO because}
we do not see how the existence and behavior of
tables and chairs and the like could be accounted for without positing a
primitive ontology---a description of matter in space and time.
The aim of a fundamental physical theory is, we believe, to describe the
world around us, and in so doing to explain our experiences to the extent
of providing an account of their macroscopic counterparts, an account of
the behavior of objects in $3$-space. Thus it seems that for a fundamental
physical theory to be satisfactory, it must involve, and fundamentally be
about, ``local beables,'' and not just a beable such as the wave function,
which is non-local.
In contrast, if a law is, like Schr\"odinger's equation,
about an abstract mathematical object, like the wave function $\psi$, living in an abstract space,
like a Hilbert space, it seems necessary that the law be supplemented with further rules or axioms in
order to make contact with a description in 3-space. For example,
formulations of classical mechanics utilizing configuration space
$\RRR^{3N}$ or phase space $\RRR^{6N}$ (such as Euler--Lagrange's or
Hamilton's) are connected to a PO in 3-space (particles with trajectories)
by the definitions of configuration space and phase space.

%A further difference between Sm and S0 concerns the inevitability of the many-worlds character. 
\z{This, at least, is how the matter seems  to us. But to}
 a proponent of S0 \z{the} existence of many worlds is a direct consequence of the Schr\"odinger equation, and \z{the} very same many worlds exist, for example, in a Bohmian universe, since Bohmian mechanics uses the same wave function. Not so, however, for a proponent of Sm. In Sm the many-worlds character arises from 
% the many contributions to $m$, not from the many contributions to $\psi$. The many-worlds character depends on
 \z{the} choice of primitive ontology and the law governing it.  A different choice, such as Bohm's law \eqref{Bohm} for a particle ontology, would  retain the single-world character.

\[***\]

We are, in fact, not the first to ask about a PO in space and time for the many-worlds view. Bell \cite{BellMW} suggested as a PO for the many-worlds view that each world consists of particles with actual positions (like a classical or Bohmian world). In a genuine many-worlds theory based on this ontology,  at every time $t$, every configuration $Q\in\RRR^{3N}$ would be realized in some world, in such a way that the distribution across the ensemble of all worlds is $|\psi_t|^2$. However, Bell himself objected that the ``other'' worlds, other than the one we are in, serve no purpose and should be discarded. In his words:
\begin{quotation}
[I]t seems to me that this multiplication of universes is extravagant, and serves no real purpose in the theory, and can simply be dropped without repercussions.
\end{quotation}
It is worth noting that this objection does not apply to Sm, as there is no easy, clean, and precise way of getting rid of all but one world in Sm. Bell, however, can remove most worlds from his picture, and thus proposes the following for the one remaining world:
\begin{quotation}
instantaneous classical configurations [$Q$] are 
supposed to exist, and to be distributed \ldots\ with probability $|\psi|^2$.
But no pairing of configurations at different times, as would be 
effected by the existence of trajectories, is supposed.
\end{quotation}
It is not clear what is meant by the last sentence, given that for every time $t$ a configuration $Q(t)$ is supposed to exist. What Bell presumably had in mind is that for every time $t$ the configuration $Q(t)$ is chosen \emph{independently} with distribution $|\psi_t|^2$. Let us call this theory Sip (S for the Schr\"odinger equation, i for independent, and p for particle ontology); in \cite{AGTZ06} it was called BMW for ``Bell's version of many-worlds.'' So in Sip, the PO consists of particles, as in Bohmian or classical mechanics, but their positions vary with time in an utterly wild and discontinuous way. (Indeed, the path $t\mapsto Q(t)$ will typically not even be a measurable function.) 

Notwithstanding the step of removing most worlds, Sip still has a certain many-worlds character, which manifests itself when one considers a time interval. Within this interval, the configuration $Q(t)$ will visit all regions of configuration space in which $\psi$ is nonzero, and those regions more often that contain more of $|\psi|^2$. So in Sip, many worlds exist, not at the same time, but one after another. For example, if after a quantum measurement the wave function of the system and the apparatus is a superposition $\sum_\alpha c_\alpha \psi_\alpha$ of contributions with the apparatus pointer pointing to different outcomes $\alpha$, then the \emph{actual} outcome, the one corresponding to the positions of the particles constituting the pointer, will be different at different times, and more often be a value with greater weight $|c_\alpha|^2$. Taking into account the occasions in the past at which wave packets split into several ones, we are led to conclude that there are also moments in time within every second, according to Sip, in which dinosaurs are still around.

Against Sip, Bell objects that the history of our world, according to Sip, is unbelievably eccentric, implying that our memories are completely unreliable, as the past was nothing like the way we remember it. It is worth noting that also this objection does not apply to Sm, as the history of every single world in Sm is much like the way we normally think of the history of our world.

Sip is related to Nelson's stochastic mechanics \cite{stochmech1,stochmech2} (in the variant due to Davidson \cite{Dav79} with arbitrary diffusion constant); in fact it can be regarded as the limiting case of stochastic mechanics in which the diffusion constant tends to infinity. As another drawback of Sip, chances seem low that this theory could ever be made relativistic, given that it relies explicitly on the concept of simultaneity. 

\[***\]

Another comparison we should make is between Sm and GRWm, the Ghirardi--Rimini--Weber (GRW) theory of spontaneous wave function collapse \cite{GRW86,Bell87} in the version with a matter density ontology \cite{BGG95,Gol98}. GRWm shares with Sm the law \eqref{mdef} for $m$, but uses a stochastic and nonlinear modification of the Schr\"odinger equation, according to which macroscopic superpositions like Schr\"odinger's cat spontaneously ``collapse'' within a fraction of a second into one contribution or another, with probabilities very close to those prescribed by the quantum formalism. As a consequence, only one of the wave packets corresponding to different ``worlds'' remains large while the others fade away, and the $m$ function of GRWm is essentially just one of the $m_\ell$ contributing to $m$ in Sm. Thus, it seems reasonable to say that GRWm does not share the many-worlds character of Sm. (On the other hand, one might argue that even in GRWm, other contributions $m_k$, $k\neq \ell$, still exist, however small they may be. Note, though, that those contributions are not just reduced in size by the GRW collapses, but also distorted, due to a large relative gradient of the tails of the Gaussian involved in the collapse, so that their evolution is  very much disturbed \cite{WM06}.)

\section{Nonlocality}\label{sec:locality}

Bell's theorem seems to show that every theory that agrees with the quantum formalism must be nonlocal \cite{Bell64}. But Bell's argument relies on the assumption that experiments have unambiguous outcomes. That is a very normal kind of assumption, but one that is inappropriate in theories with a many-worlds character, as Bell concedes in the passage quoted in the beginning of this article. Because of its ontological clarity, Sm provides an occasion to analyze the relevance of the many-worlds character to locality. So is Sm a local theory or not? 

We first observe that in the absence of interaction between two disjoint regions $A$ and $B$ of space, experimenters in $A$ have no way of influencing $m|_B$, the matter density in $B$. After all, if
\[
\psi=\psi(q_1,\ldots,q_{M},r_1,\ldots,r_{N})=\psi(\vec{q},\vec{r})
\]
is a wave function for which some variables are confined to $A$, $q_1,\ldots,q_{M} \in A$, and some to $B$, $r_1,\ldots, r_{N}\in B$, \n{then} $m|_B$ depends on $\psi$ only through the reduced density matrix associated with $B$, $\rho_B = \tr_A |\psi\rangle \langle \psi|$, where $\tr_A$ means the partial trace over the variables $\vec{q}=(q_1,\ldots,q_{M})$. Indeed, 
\begin{equation}
\rho_B(\vec{r};\vec{s}) = 
\int d^{3M}\vec{q} \:\: \psi^*(\vec{q},\vec{r}) \, \psi(\vec{q},\vec{s})
\end{equation}
and, for $x\in B$,
\begin{equation}\label{mrho}
m(x) = \sum_{j=1}^{N} m_{M+j} \int d^{3N}\vec{r} \:\: \delta^3(x-r_j)\:
\rho_B(\vec{r};\vec{r})\,.
\end{equation}
\n{Since, as is well known, $\rho_B$ will}   not depend on any external fields at work in $A$, fields that the experimenters may have set up to influence the matter governed by $\psi$, \n{for} as long as there is no interaction between $A$ and $B$, \n{it follows that the same thing is true of $m|_B$}. This shows that experimenters in $A$ cannot influence $m|_B$.

And yet, Sm is nonlocal. To see this, consider an Einstein--Podolsky--Rosen (EPR) experiment, starting with two electrons in the singlet state, one in Alice's lab $A$ and the other in Bob's lab $B$. While there is no interaction between $A$ and $B$, Alice and Bob each perform a Stern--Gerlach experiment in the $z$ direction. \n{Now} consider a time $t$ just after detectors have clicked on both sides. Recall that in ordinary quantum mechanics the outcome has probability $\tfrac{1}{2}$ to be (up, down) and probability $\tfrac{1}{2}$ to be (down, up). Hence, in Sm the wave function $\psi=\psi_t$ of the EPR pair together with the detectors (and other \n{devices}) splits \n{into} two macroscopically disjoint packets,
\begin{equation}\label{zzpsi}
\psi=\sum_\ell \psi_\ell = \psi_\mathrm{(up, down)}+\psi_\mathrm{(down, up)}\,,
\end{equation}
and correspondingly,
\begin{equation}\label{zzm}
m= \sum_\ell m_\ell = m_\mathrm{(up, down)} + m_\mathrm{(down, up)}\,.
\end{equation}
Thus the world in which Alice's result is ``up'' is the same world as the one in which Bob's result is ``down,'' and it is this fact that is created in a nonlocal way.

To connect, and contrast, this nonlocality with the fact that Alice cannot influence $m|_B$, we note that the $m$ function alone, while revealing that there are two worlds in $A$ (corresponding to the results ``up'' and ``down'') and two worlds in $B$ (corresponding to the results ``up'' and ``down''), does not encode the information conveying which world in $A$ is the same as which world in $B$. That is, the \emph{pairing of worlds} cannot be read off from $m(\cdot, t)$ even though it is an objective fact of Sm at time $t$, defined by means of the wave function $\psi_t$. 

Moreover, even though Alice cannot influence the PO in $B$, she can influence other physical facts pertaining to $B$ as follows. Consider now two options for Alice: she can carry out a Stern--Gerlach experiment in either the $z$ direction or the $x$ direction (what is often called measuring $\sigma_z$ or $\sigma_x$). Suppose further that $t_1$ is a time at which a detector in $A$ has already clicked but the electron in $B$ has not yet reached its Stern--Gerlach magnet. Then the wave function $\psi=\psi_{t_1}$ of the EPR pair and the detectors in $A$ together is either---if Alice chose the $z$ direction---of the form
\begin{align}
\psi &= \nonumber
\tfrac{1}{\sqrt{2}}  \: \bigl|\uparrow,z=+1\bigr\rangle_A \:
\bigl|\downarrow,z=0 \bigr\rangle_B \: \bigl|\text{``up''}\bigr\rangle\\
&+\tfrac{1}{\sqrt{2}}  \: \bigl|\downarrow,z=-1\bigr\rangle_A \:
\bigl| \uparrow,z=0 \bigr\rangle_B \: \bigl|\text{``down''}\bigr\rangle
\end{align}
(with the first two factors referring to spin and position of the EPR pair and the third to the detectors in $A$), or---if Alice chose the $x$ direction---of the form
\begin{align}
\psi &= \nonumber
\tfrac{1}{\sqrt{2}}  \: \bigl|\rightarrow,x=+1\bigr\rangle_A \: 
\bigl|\leftarrow,z=0\bigr\rangle_B 
\: \bigl|\text{``right''}\bigr\rangle \\%_\text{detectors}
&+\tfrac{1}{\sqrt{2}}  \: \bigl|\leftarrow,x=-1\bigr\rangle_A  \: 
\bigl| \rightarrow,z=0 \bigr\rangle_B 
\: \bigl|\text{``left''}\bigr\rangle \,.
\end{align}
Now suppose that at time $t_2>t_1$, the electron in Bob's lab has passed through a Stern--Gerlach magnetic field oriented in the $z$ direction, but not yet reached the detectors. Then the above expressions for $\psi=\psi_{t_1}$ have to be modified as follows for $\psi=\psi_{t_2}$ (of the EPR pair and the detectors in $A$): using 
\begin{equation}
\bigl|\rightarrow\bigr\rangle = 
\tfrac{1}{\sqrt{2}}\bigl(\bigl|\uparrow\bigr\rangle + \bigl|\downarrow\bigr\rangle\bigr)
\quad \text{and} \quad 
\bigl|\leftarrow\bigr\rangle = 
\tfrac{1}{\sqrt{2}} \bigl(\bigl|\uparrow\bigr\rangle - \bigl|\downarrow\bigr\rangle\bigr)\,,
\end{equation}
we have that either---if Alice chose the $z$ direction---
\begin{align}
\psi &= \nonumber
\tfrac{1}{\sqrt{2}} \: \bigl|\uparrow,z=+1\bigr\rangle_A \:
\bigl|\downarrow,z=-1 \bigr\rangle_B \: \bigl|\text{``up''}\bigr\rangle \\%_\text{detectors}
&+\tfrac{1}{\sqrt{2}}  \: \bigl|\downarrow,z=-1\bigr\rangle_A \:
\bigl| \uparrow,z=+1 \bigr\rangle_B \: \bigl|\text{``down''}\bigr\rangle 
\end{align}
or---if Alice chose the $x$ direction---
\begin{align}
\psi& = \nonumber
\tfrac{1}{2} \: \bigl|\rightarrow,x=+1\bigr\rangle_A \: 
\bigl(\bigl|\uparrow,z=+1\bigr\rangle_B - \bigl|\downarrow,z=-1\bigr\rangle_B\bigr)
\: \bigl|\text{``right''}\bigr\rangle \\%_\text{detectors}
&+ \tfrac{1}{2} \: \bigl|\leftarrow,x=-1\bigr\rangle_A \: 
\bigl(\bigl| \uparrow,z=+1 \bigr\rangle_B + \bigl|\downarrow,z=-1\bigr\rangle_B\bigr)
\: \bigl|\text{``left''}\bigr\rangle\,.
\end{align}
As a consequence, at time $t_2$ the decomposition $m=\sum m_\ell = m_1+m_2$ into worlds reads, on the $B$ side, either---if Alice chose the $z$ direction---
\begin{equation}
m_1|_B = \tfrac{1}{2} m_{z=-1}\,,\quad m_2|_B = \tfrac{1}{2} m_{z=+1} 
\end{equation}
(with $m_{z=-1}$ a unit bump centered at $z=-1$, etc.)
or---if Alice chose the $x$ direction---
\begin{equation}
m_1|_B = \tfrac{1}{4} m_{z=-1} + \tfrac{1}{4} m_{z=+1}\,,\quad 
m_2|_B = \tfrac{1}{4} m_{z=-1} + \tfrac{1}{4} m_{z=+1}\,. 
\end{equation}
That is, while $m(x)$ for $x\in B$ is unaffected by Alice's choice, each $m_\ell(x)$ is affected. This is an example of an objective fact pertaining to region $B$ that is influenced by Alice's choice, and illustrates that the nonlocality of Sm is even of the kind involving instantaneous influences.\footnote{It is interesting that Sm turns out to be nonlocal in situations that do not seem to require nonlocality in a single-world framework: If both Alice and Bob choose the $z$ direction, then the correlation can of course be explained locally by means of a ``hidden variable'' (that was the point of Einstein, Podolsky and Rosen \cite{EPR35}). If Alice chooses the $x$ and Bob the $z$ direction then the outcomes of both sides are independent. So, if Alice can only choose between $z$ and $x$ then the statistics can be explained by two independent hidden variables associated with the $z$ and $x$ directions.}

The situation of nonlocality in Sm can be compared to that in the ``many minds'' picture described by Albert and Loewer \cite{AL88}. There, the PO is replaced by a collection of purely mental events; Alice has many minds, some of which see ``up'' and some ``down,'' and so does Bob, but no pairing is assumed that would specify which of Bob's minds is in the same world as which of Alice's. This parallels the absence of pairing between Bob's worlds and Alice's worlds in the $m$ function. The clarity of Sm helps exemplify that this fact alone does not imply locality: Even if one assumes the absence of a pairing in the PO (or, for ``many minds,'' in the mental events replacing the PO), the non-primitive ontology (i.e., the wave function) may define such a pairing nonetheless. For the same reason, also ``many minds'' should be regarded as nonlocal. Of course, the nonlocality of Sm is already suggested by the facts that Sm cannot be formulated purely in terms of local variables (but needs the nonlocal variable $\psi$), and that EPR (or Bell) correlations in Sm do not propagate through space at finite speed.

\section{Relativity}\label{sec:relativity}

Even though Sm is nonlocal, it can easily be made relativistic, \z{at least formally, neglecting cut-offs and renormalization. We acknowledge  that  to go beyond  a formal theory such as sketched below to one that is well defined and physically adequate would of course be a formidable challenge.} 

For any relativistic quantum theory, consider the Heisenberg picture with fixed state vector $\psi$, let $T_{\mu\nu}(x,t)$ be the stress-energy tensor operator for the space-time point $(x,t)$, and set
\begin{equation}\label{relmdef}
m_{\mu\nu}(x,t) = \scp{\psi}{T_{\mu\nu}(x,t)|\psi}\,.
\end{equation}
This tensor field on space-time is arguably the most obvious relativistic analog of the formula \eqref{mdef}, for what could in ordinary quantum mechanics be called the average mass distribution. Indeed, in the nonrelativistic limit, $m_{\mu\nu}(x,t)$ should have time-time component
\begin{equation}
m_{00}(x,t)=m(x,t) \, c^2
\end{equation}
with $m(x,t)$ as in \eqref{mdef}, and all other components negligible. The theory with PO given by \eqref{relmdef} is relativistically invariant because of the relativistic invariance of the underlying quantum theory and that of the operator-valued tensor field $T_{\mu\nu}(x,t)$.

Other relativistic laws than \eqref{relmdef} are conceivable. In fact, the concept of matter density per se does not even select whether the relativistic analog of the $m(\cdot)$ function should be a scalar, vector, or tensor field; the above choice of tensor field was inspired by the relativistic concept of mass-energy, but, as mentioned before, the matter density function need not be linked to masses.

\section{Probability}\label{sec:probability}

In ordinary quantum mechanics, the outcome of a ``quantum measurement'' (say, a Stern--Gerlach experiment) is regarded as random with certain probabilities. In Sm, though, all possible outcomes are realized in different worlds, so it is not obvious how it can make sense to talk of probabilities at all. What are these probabilities the probabilities \emph{of}?

This problem is often called the ``incoherence problem,'' applying to any many-worlds interpretation of quantum mechanics, be it in the form originally put forward by Everett \cite{evethesis, Eve57} or in other formulations (see, e.g., \cite{deutsch0}). Moreover, most authors agree that once this problem is solved, there is still the quantitative problem of showing that the probabilities agree with the quantum-mechanical ones; see, e.g., \cite{greaves2004}. 
In recent years various proposals have been put forward to solve these problems. David Deutsch has suggested that probabilities can be understood in terms of rational action \cite{deutsch} and that one should prove, via decision theory, that a ``rational'' agent  who believes himself to be in a many-worlds universe, should nevertheless make decisions {\em as if}  the quantum probabilities gave the  chances for the results of experiments in the usual way; in this regard, see also the contribution of David Wallace \cite{wallace}. Lev Vaidman has put forward his ``sleeping pill'' argument to support the validity of the ignorance interpretation of probability \cite{vaidman, vaidman1}. Recently, Simon Saunders and Wallace  have considered a ``semantic turn'' in order to ensure the truth of utterances typically made about quantum mechanical contingencies, including statements of uncertainty, by speakers living in a many-worlds universe \cite{wallace06, saunderswallace}. See also \cite{tappenden, greaves2004, lewis2007, baker2007}.

Since Sm is a many-worlds formulation of quantum mechanics---albeit with a precise primitive ontology---any of the proposals mentioned above about the meaning of probabilities in a many-worlds setting can equally well be considered in Sm. We prefer, however, Everett's approach \cite{evethesis, Eve57}, that of \x{denying that the incoherence problem is a genuine problem (for more on this see Section~\ref{sec:uncertainty}) and appealing to {\em typicality} for the quantitative problem. Typicality is} a notion that goes back at least to Ludwig Boltzmann's mechanical analysis of the  second law of thermodynamics \cite{golsteinstat}, and that, in recent years, has been used for explaining the emergence of quantum randomness in Bohmian mechanics \cite{DGZ92}.\footnote{In  Bohmian mechanics  different histories of the world, corresponding to different initial configurations, are possible for the same wave function $\Psi$ of the universe, and the observed frequencies may agree with quantum mechanics in some histories but not in others. Thus a concept of ``typical history'' is needed. The only known candidate for this concept that is time translation invariant is the one given by the $|\Psi|^2$ measure. This measure is {\em equivariant} \cite{DGZ92}, a property which expresses the mutual compatibility of the Schr\"odinger evolution of the wave function and the Bohmian motion of the configuration.  This measure is used in the following way:
A property $P$ is typical if it holds true for the
overwhelming majority of histories $Q(t)$ of a Bohmian universe. More
precisely, suppose that $\Psi_t$ is the wave function of a universe
governed by Bohmian mechanics; a property $P$, which a solution $Q(t)$ of
the guiding equation for the entire universe can have or not have, is
called \emph{typical} if the set $S_0(P)$ of all initial configurations
$Q(0)$ leading to a history $Q(t)$ with the property $P$ has size very
close to one,
\begin{equation}\label{typ}
  \int_{S_0(P)} |\Psi_0 (q)|^2 dq= 1 - \varepsilon\,, \quad 0 \leq
  \varepsilon \ll 1\,,
\end{equation}
with ``size'' understood relative to the $|\Psi_0|^2$ distribution on the
configuration space of the universe.  For instance, think of $P$ as the
property that a particular sequence of experiments yields results that look
random (accepted by a suitable statistical test), governed by the
appropriate quantum distribution.  One can show, using the {\em law of large
numbers}, that $P$ is typical; see \cite{DGZ92} for a thorough
discussion.}  
According to Everett \cite{Eve57}: 

\begin{quote}
We wish to make quantitative statements about the relative frequencies
of the different possible results \ldots\ for a typical observer state;
but to accomplish this we must have a method for selecting a typical
element from a superposition of orthogonal states. \ldots

The situation here is fully analogous to that of classical statistical 
mechanics, where one puts a measure on trajectories of systems in the 
phase space by placing a measure on the phase space itself, and then 
making assertions which hold for ``almost all'' trajectories (such as 
ergodicity, quasi-ergodicity, etc).  This notion of ``almost all'' depends 
here also upon the choice of measure, which is in this case taken to be 
Lebesgue measure on the phase space. \ldots  [T]he choice of Lebesgue measure on the phase space can be justified by the fact that it is the only choice for which the ``conservation of probability'' holds, (Liouville's theorem) and hence the only choice which makes possible any reasonable statistical deductions at all. 
 
In our case, we wish to make statements about ``trajectories'' of observers. However, for us a trajectory is constantly branching (transforming from state to superposition) with each successive measurement.
\end{quote}

Let us explain how that works in Sm. It is useful to focus on the following statement: 
\begin{equation}
\mbox{
\begin{minipage}{0.85\textwidth}
\textit{The relative frequencies for the results of experiments that a typical observer sees agree, within appropriate limits, with the probabilities specified by the quantum formalism.}
\end{minipage}}
\label{stat:one}
\end{equation}
We elaborate on this statement below. The idea is that a derivation of this statement amounts to a justification of our use of the quantum probabilities. For a discussion of the idea of a \emph{typical observer} in a different context see \cite{Gott93}, \cite[Chap.~5]{Gott01}, where the rule that we humans should see what a typical observer sees was called the ``Copernican principle.''

By what a ``typical observer sees,'' be it relative frequencies or any other sort of behavior corresponding to some property $P$, we mean that $P$ occurs in ``most'' worlds. When this is true, we often also say that the behavior is typical, or that $P$ typically holds, or that $P$ is typical. It is, of course, crucial here to specify exactly what is meant by ``most''---what is meant by saying that $P$ is typical.

The sense of typical we have in mind is given by assigning to each world $m_l$ a weight
\begin{equation}\label{weight}
\mu_{\ell} = \int d^3x \, m_\ell(x,t)\,.
\end{equation}
We say that 
\begin{equation}
\mbox{
\begin{minipage}{0.85\textwidth}
  \emph{A property $P$ holds typically (or, for most worlds) 
  if and only if the sum of the weights $\mu_{\ell}$, given by (\ref{weight}),  
  of those worlds for which $P$ holds is 
  very near the sum of the weights of all worlds.}
\end{minipage}}
\label{stat:two}
\end{equation}
In the next section we will discuss why we believe that this is a reasonable notion of typicality, one such that we should expect to see what is typical. Let us now explore its mathematical consequences.\footnote{One might be tempted to think that some of the worlds, those represented with less weight $\mu_{\ell}$,
are somehow less real, corresponding perhaps to a lesser degree of existence (a ``measure of existence'' was considered by Vaidman \cite{vaidman1}). But we do not think that there can be different degrees of existence, and we certainly see no basis for such a position in Sm.}

In terms of the decomposition \eqref{psiell} of the wave function $\psi$ into macroscopically different contributions $\psi_\ell$, the weight can be expressed as follows, according to the definition \eqref{mdef} of the $m$ function:
\begin{equation}\label{weightell}
\mu_{\ell} = \int d^3x \, m_\ell(x,t) = \|\psi_\ell\|^2 \sum_{i=1}^N m_i
\end{equation}
(recall that $m_i$ are the mass parameters associated with the $N$ ``particles'').
That is, the weights we associate with different worlds are the same weights, up to a factor $\sum m_i$, as would usually be associated with different worlds in a many-worlds framework. We can also rephrase the typicality of $P$ in terms of the $\psi_\ell$: Let $\Lset$ be the set of all indices $\ell$, $\Lset=\{1,\ldots,\Lnum\}$, and $\Lset(P)$ the set of those indices $\ell$ such that the world with index $\ell$ has the property $P$. By \eqref{stat:two}, the property $P$ holds typically if and only if
\begin{equation}\label{typmw}
\frac{\sum\limits_{\ell\in \Lset(P)}  \mu_{\ell}}
{\sum\limits_{\ell \in \Lset}  \, \mu_{\ell} } = 1-\varepsilon\,, 
\quad 0\leq \varepsilon \ll 1\,.
\end{equation}
Since the $\psi_\ell$ do not overlap, and assuming $\|\psi\|=1$ as usual, we have that
\[
\sum_{\ell \in \Lset} \mu_{\ell}  = \|\psi\|^2 \sum_i m_i = \sum_i m_i\,.
\]
Thus, $P$ holds typically if and only if\footnote{Note the similarity  between (\ref{typ}) and (\ref{typmw1}).}
\begin{equation}
\sum_{\ell\in \Lset(P)} \|\psi_\ell\|^2 = 1-\varepsilon\,,
\quad 0\leq \varepsilon \ll 1\,.
\label{typmw1}
\end{equation}

Everett showed that with the sense of typicality provided by the weights (\ref{weight},\ref{weightell}) the law of large numbers yields \eqref{stat:one}. A simple example should suffice here. Consider an observer performing a large number $\num$ of independent Stern--Gerlach experiments for which quantum mechanics predicts ``spin up'' with probability $p$ and ``spin down'' with probability $q=1-p$. Let this $\num$-part experiment begin at time $t_0$ and end at time $t$; let us focus on just one world at time $t_0$. Assume that the sequence of outcomes, such as
\begin{equation}\label{sequence}
\uparrow \downarrow\downarrow\uparrow\ldots\downarrow \uparrow\uparrow\uparrow\,,
\end{equation}
gets recorded macroscopically, and thus in $m_\ell(\cdot,t)$. The one world at \n{time} $t_0$  splits into $\Lnum\geq 2^\num$ worlds at \n{time} $t$,
\begin{equation}\label{branching}
\psi=\psi(t) = \sum_{\ell=1}^\Lnum \psi_\ell(t)\,,\qquad
m(x,t) = \sum_{\ell=1}^\Lnum m_\ell(x,t)\,.
\end{equation}
Now some of the worlds at time $t$ feature a sequence in which the relative frequencies of the outcomes agree, within appropriate limits, with the quantum probabilities $p$ and $q$.  However, this is true only of \emph{some} worlds, but not all. It is a property $P$ that a world may have or not have. 

Is $P$ typical? Let $\Lset(k)$ be the set of those $\ell$ such that the world $m_\ell$ features a sequence of $k$ spins up and $\num-k$ spins down; taken together, these worlds have weight
\begin{equation}
\sum_{\ell\in\Lset(k)}\mu_\ell = 
\Bigl(\sum_i m_i\Bigr)\sum_{\ell\in\Lset(k)} \|\psi_\ell\|^2= 
\Bigl(\sum_i m_i\Bigr) \binom{\num}{k} p^k q^{\num-k}\,.
\end{equation}
Since $\num$ is large, the weight is overwhelmingly concentrated on those worlds for which the relative frequency $k/\num$ of ``up'' is close to $p$. This follows from the law of large numbers, which ensures that, if we generated a sequence of $\num$ independent random outcomes, each ``up'' with probability $p$  or ``down'' with probability $q$, then the relative frequency of ``up'' will be close to $p$ with probability close to 1. Thus the total weight of the worlds with $k/\num\approx p$ is close to the total weight. This illustrates how \eqref{stat:two} yields \eqref{stat:one}. The upshot is that Sm is empirically equivalent to both orthodox quantum mechanics and Bohmian mechanics.\footnote{We note the following subtlety about the empirical equivalence between Sm and Bohmian mechanics. Even though there is no experiment that could distinguish them, there exist contrived situations in which Sm does not make the same empirical prediction as Bohmian mechanics, but rather makes no empirical prediction at all. Namely, there exist contrived situations in which Sm provides no recognizable macroscopic objects, while any experimental test would, of course, require the existence of \x{such} objects, in particular of pointers to register the outcome and of humans or other beings as experimenters. For example, suppose that 3-space is not $\RRR^3$ but a 3-torus $(S^1)^3$, where $S^1$ denotes a circle (of some large perimeter). Then some wave functions on configuration space $(S^1)^{3N}$ are invariant under translations of $(S^1)^3$. Such wave functions can be obtained from any $\psi$ by superposing all translates of $\psi$. They would appear completely acceptable in Bohmian mechanics but would lead to a profoundly problematical state of the PO in Sm, namely a constant $m$ function. To see this, note that if two wave functions are translates of each other then the $m$ functions they give rise to are translates of each other as well; as a consequence, a translation invariant wave function $\psi$ (which may be very nontrivial) gives rise to a translation invariant $m$ function (which must be constant). Of course, this fact is not fatal to the viability of Sm, as  the wave function of the universe need not  be translation invariant.}

In both Sm and Bohmian mechanics, typicality is used for two purposes: prediction and explanation. Namely, when deriving predictions from Bohmian mechanics we claim that the typical behavior will occur, even if there are possible universes in which different behavior occurs. For explanation of why the world looks the way it does, we say that it is typical for a Bohmian world to look that way. Likewise in Sm: When we want to make predictions, we know that a property like $P$ will hold in some worlds $m_\ell$ and not in others, so what we predict is the typical behavior---the one that occurs in most worlds (with the weighted notion of ``most''); and the explanation for why we see a certain behavior is that it occurs in most worlds. Insofar as the typicality reasoning is concerned, in Sm the \emph{world we are in} plays the same role as the \emph{actual world} in Bohmian mechanics.\footnote{What is different about the use of typicality in the two theories is that while in Bohmian mechanics typicality is used for explaining \emph{physical} facts, in Sm it is used for explaining \emph{indexical} facts. Indexical statements are statements referring to concepts like ``here,'' ``now,'' or ``I.'' A simple example of an indexical statement is ``there are five coins in my pocket.'' In physics, once I am told all physical facts about my universe, I may still need to be told where I am in this picture: which space-time location corresponds to here-now, and furthermore, for any theory with many-worlds character, in which of the worlds to find \emph{me}. Those are indexical facts. The indexical fact to be explained here is that I find myself in a world with property $P$.}

\section{Typicality}
\label{sec:typicality}

In this section, we address the following two questions: Should not the concept of typicality (or that of ``most'' worlds) be based on the \emph{number} of worlds, disregarding the weights $\mu_\ell$? And, which reasons select \eqref{weight} as the rule for determining the weights?

It would seem that counting would provide a better measure of typicality, maybe even the only acceptable one. And counting would also seem to lead to rather different predictions. After all, in the example above involving $\num\gg 1$ Stern--Gerlach experiments, if the worlds are taken to be in one-to-one correspondence with the possible outcomes (i.e., the sequences of ups and downs) then, by the law of large numbers, the worlds in which the relative frequency $k/\num$ of ``up'' is approximately $1/2$ far outnumber those in which $k/\num\approx p$ (provided $p$ is sufficiently different from $1/2$).

But counting worlds is not well defined; there is no fact of the matter as to how many worlds have some property $P$. In this respect, ``worlds'' are not like beans (that can be counted) but more like clouds. The decomposition $\psi = \sum \psi_\ell$ \z{is associated with} an orthogonal decomposition $\Hilbert=\oplus_\ell \Hilbert_\ell$ of the Hilbert space $\Hilbert$ into subspaces $\Hilbert_\ell$ corresponding to different macrostates \cite{vN}, a decomposition that is inevitably arbitrary, due to the vagueness of the notion of ``macroscopic'' or, in other words, due to the arbitrariness of the boundaries between macrostates.\footnote{\z{The suggestion that world count is ill-defined has been  discussed recently in  \cite{wallace}.}} Concretely, it is often unclear whether two wave packets $\phi_1$ and $\phi_2$ should be regarded as ``macroscopically different'' or not; as a consequence, it is then unclear whether $\psi = \phi_1+\phi_2$ should be regarded as two worlds, $\psi_1=\phi_1$ and $\psi_2=\phi_2$, or as one world, $\psi_1=\phi_1+\phi_2$. Indeed, the decomposition of $\psi$ into ``its macroscopically different contributions'' $\psi_\ell$ will \emph{usually} depend on our interpretation of ``macroscopically different.'' For example, if $\psi$ as a function of the center-of-mass coordinate of a meter pointer is smeared out, into how many ``macroscopically different'' parts should we divide it? 
And in the example above of $\num$ Stern--Gerlach experiments, should we regard the worlds as corresponding to different outcomes (given as sequences of ups and downs), or should we choose a finer decomposition by taking into account the times when detectors clicked, and regard contributions to $\psi$ that correspond to different times (but the same sequence of ups and downs) as different worlds? For many purposes, the ambiguity inherent in the notion ``macroscopically different'' is not a problem, but for the purpose of counting worlds it is.

However, if we use the weights $\mu_\ell$ (or, equivalently, $\|\psi_\ell\|^2$) then, while there is still the same amount of arbitrariness in the decomposition $\psi=\sum\psi_\ell$, the weight associated with the property $P$, 
\[
\mu(P)=\sum_{\ell\in\Lset(P)}\mu_\ell\,,
\]
is unambiguous, as a consequence of what Everett \cite{evethesis,Eve57} called the \emph{additivity} of the weights: When we further decompose a contribution $\psi_\ell$ into $\sum_{\ell'} \psi_{\ell,\ell'}$ then the norm squares add according to
\begin{equation}
\|\psi_\ell\|^2 = \sum_{\ell'} \|\psi_{\ell,\ell'}\|^2\,.
\end{equation}
This follows if the $\psi_{\ell,\ell'}$ have disjoint supports in configuration space. (In fact, Everett showed that the weights must be equal, up to an overall factor, to $\|\psi_\ell\|^2$ if they are given by some fixed function $f(\|\psi_\ell\|)$ and additive, $f(\|\psi_\ell\|)= \sum_{\ell'} f(\|\psi_{\ell,\ell'}\|)$, where $\psi_{\ell,\ell'}$ are mutually orthogonal. However, he did not make explicit the connection between additivity and the ambiguity of the notion of the  macroscopically different.)

Thus, counting the worlds is not an option. But even in theories in which the concept of world is precisely defined and thus allows us to count worlds, weights may arise naturally in the form of multiplicities, associated with representing several worlds with the same configuration by one world with multiplicity.

Another factor supporting the use of the weights $\mu_\ell$ \eqref{weight} for the  measure of typicality is their \emph{quasi-equivariance}, i.e., the two facts, analogous to the equivariance of the $|\psi|^2$ distribution in Bohmian mechanics, that the weight $\mu_\ell$ of a world does not change under the unitary time evolution unless it splits, and that when a world splits then the sum of the weights after splitting is the same as the weight before splitting. The former is a consequence of unitarity and the fact that $\mu_\ell$ is proportional to $\|\psi_\ell\|^2$, and the latter follows from the additivity mentioned above. Quasi-equivariance is relevant since, as Everett says in the passage quoted above, ``we wish to make statements about ``trajectories'' of observers.'' As a consequence of quasi-equivariance, memories and records obey the following type of consistency: If it is typical at time $t_1$ that, say, between $33\%$ and $34\%$ of the outcomes of a certain experiment are ``up'' then it is typical at time $t_2>t_1$ that between $33\%$ and $34\%$ of the records of those outcomes are records of ``up.'' 

The notions of typicality and quasi-equivariance we have considered are just what Everett considered after the passage quoted above:
\begin{quote}
To have a requirement analogous to the ``conservation of probability'' in the 
classical case, we demand that the measure assigned to a trajectory at 
one time shall equal the sum of the measures of its separate branches at 
a later time. This is precisely the additivity requirement which we imposed and which leads uniquely to the choice of square-amplitude measure. Our procedure is therefore quite as justified as that of classical statistical mechanics. 
\end{quote}
In other words, Everett's assessment is that the most natural measure is indeed the one using $\|\psi_\ell\|^2$ weights, and we agree. In addition, owing to Sm's greater ontological clarity, we believe that  Everett's analysis, when applied to Sm, becomes even more transparent and compelling than for more standard versions of the many-worlds interpretation.

\section{Uncertainty}
\label{sec:uncertainty}

Probabilities are often regarded as expressions of our lack of knowledge, of ignorance and uncertainty; the typicality approach, however, does not directly involve uncertainty. So let us make some remarks about the status of uncertainty \x{in Sm}.

Since Bohmian mechanics is a deterministic theory, it also gives rise to the question about the meaning of probabilities. But in Bohmian mechanics this question is much less problematical than in Sm; in a Stern--Gerlach experiment, for example, the outcome depends on the initial wave function and the initial position of the particle, and while we may know the wave function, we cannot know the position with sufficient precision to infer the outcome (except when the initial wave function is an eigenstate of the relevant spin operator). But in Sm we cannot be uncertain about what ``the'' outcome of a Stern--Gerlach experiment will be, since we know that both outcomes will be realized. If we believe we are living in a many-worlds universe, we should regard our feelings of uncertainty about the future as sheer illusion.  

As shocking as this may seem, \x{however,} it should not be held against many-worlds theories. After all, modern physics has accustomed us to the illusory character of many of our experiences; for example, according to the standard understanding of relativity (at least among physicists), our feeling of the passage of time is also a sheer illusion (``however persistent,'' as Albert Einstein wrote to the widow of Michele Besso). Likewise, the fact that in Sm (as in any physical theory with a many-worlds character) there is a severe gap between metaphysics and experiences---i.e., the fact that the reality is very different from what our experiences suggest, in that the future is not as uncertain as we normally imagine and that there are other worlds we do not see---need not conflict with the goal of explaining our experiences. It seems not at all impossible to explain why observers who believe in a many-worlds universe nevertheless behave as if they were uncertain about the future; some approaches are presented in, e.g., \cite{wallace06, tappenden, greaves2004, lewis2007, baker2007, saunderswallace}. (In fact, we would presumably often feel uncertain in a many-worlds theory for the same evolutionary reasons as for a single-world theory.)  However, this problem should not be confused with the problem of explaining the origin of physical probabilities, i.e., of explaining the relative frequencies we see, a problem resolved by the typicality analysis.

\section{Summary}

We have shown that Schr\"odinger's first interpretation of quantum mechanics, in which the wave function is regarded as describing a continuous distribution of matter in space and arguably the most naively obvious interpretation of quantum mechanics, has a surprising many-worlds character. We have also shown that insofar as this theory makes any consistent predictions at all, these are the usual predictions of textbook quantum mechanics.

\bigskip

\noindent\textit{Funding.} 
This work was supported by the National Science Foundation [DMS-0504504 to S.G.]; and Istituto Nazionale di Fisica Nucleare [to N.Z.].
%S.~Goldstein is supported in part by National Science Foundation [grant DMS-0504504].
%N.~Zangh\`\i\ is supported in part by Istituto Nazionale di Fisica Nucleare. 

\bigskip

\noindent\textit{Acknowledgments.} 
We thank David Albert (Columbia University), Cian Dorr (Oxford), Detlef D\"urr (LMU M\"unchen), Adam Elga (Princeton), Ned Hall (Harvard), Barry Loewer (Rutgers), Tim Maudlin (Rutgers), Travis Norsen (Marlboro), Daniel Victor Tausk (S\~ao Paulo), \z{David Wallace (Oxford),} and Hans Westman (Sydney) for useful discussions and comments on previous versions of this paper. We are particularly grateful to Travis Norsen for discussions on nonlocality. %The work of S. Goldstein was supported in part by NSF Grant DMS-0504504 and that of Nino Zangh\`\i\   by INFN. 

\end{document}